\documentclass[aps,eprint,amsmath,amssymb,superscriptaddress,showpacs,prl]{revtex4-1}
\usepackage{graphicx}
\usepackage{epsfig}
\usepackage{bm}
\usepackage{amsmath}
\usepackage{CJK}
\usepackage{fancyhdr}
\usepackage{caption2}
\usepackage{subfigure}
\usepackage{color}

\begin{document}
\begin{CJK*}{GBK}{kai}
\title{A new ignition scheme using hybrid indirect-direct drive for inertial confinement fusion}
\author{Zhengfeng Fan}
\email{zffan@yahoo.com.cn} \affiliation{Institute of Applied Physics
and Computational Mathematics, Beijing 100088, China}
\author{Mo Chen}
\affiliation{Institute of Applied Physics and Computational
Mathematics, Beijing 100088, China}
\author{Zhensheng Dai}
\affiliation{Institute of Applied Physics and Computational
Mathematics, Beijing 100088, China}
\author{Hong-bo Cai}
\affiliation{Institute of Applied Physics and Computational
Mathematics, Beijing 100088, China}
\author{Shao-ping Zhu}
\affiliation{Institute of Applied Physics and Computational
Mathematics, Beijing 100088, China}
\author{W. Y. Zhang}
\affiliation{Institute of Applied Physics and Computational
Mathematics, Beijing 100088, China}
\author{X. T. He}
\email[]{xthe@iapcm.ac.cn} \affiliation{Institute of Applied Physics
and Computational Mathematics, Beijing 100088, China}
\affiliation{Center for Applied Physics and Technology, Peking
University, Beijing 100871, China}

\begin{abstract}
A new hybrid indirect-direct-drive ignition scheme is proposed for
inertial confinement fusion: a cryogenic capsule encased in a
hohlraum is first compressed symmetrically by indirect-drive x-rays,
and then accelerated and ignited by both direct-drive lasers and
x-rays. A steady high-density plateau newly formed between the
radiation and electron ablation fronts suppresses the rarefaction at
the radiation ablation front and greatly enhances the drive
pressure. Meanwhile, multiple shock reflections at the fuel/hot-spot
interface are prevented during capsule deceleration. Thus rapid
ignition and burn are realized. In comparison with the conventional
indirect drive, the hybrid drive implodes the capsule with a higher
velocity ($\sim$4.3 $\times$ $10^7$ cm/s) and a much lower
convergence ratio ($\sim$25), and the growth of hydrodynamic
instabilities is significantly reduced, especially at the
fuel/hot-spot interface.
\end{abstract}

\pacs{52.57.-z, 28.52.Cx}

\maketitle

In central hot spot ignition scheme of inertial confinement fusion
(ICF) \cite{Nuckolls_1972,Lindl_1998,Atzeni_2004}, a spherical
capsule which is composed of deuterium-tritium (DT) gas, DT fuel and
an ablator, is cryogenically imploded to a high velocity, then the
DT fuel is highly compressed and a hot spot is formed at the capsule
center due to spherical convergent compression effect. When the
alpha-particle heating of the hot spot exceeds the total energy
losses, central ignition occurs and a burn wave propagates into the
surrounding cold fuel. Two main implosion schemes, direct drive
\cite{Bodner_1998} and indirect drive \cite{Lindl_2004}, have been
proposed for realizing central hot spot ignition. The National
Ignition Campaign (NIC) \cite{Moses_2005} experiments have made
great progresses towards ignition using indirect-drive targets. For
example, the National Ignition Facility is now capable of delivering
1.9 MJ of 0.35-$\mu$m laser light at 500 TW, and the assembled fuel
areal density has reached about 1.3 g/cm$^2$ which is 80\% of the
ignition goal \cite{Mackinnon_2012,Landen_2012}. However, the NIC
experiments are facing challenges, and the experimental ignition
threshold factor which is a metric of the progress towards ignition
needs to be increased by an order of magnitude to meet the ignition
requirement. Besides laser plasma instabilities (LPI), two major
issues in implosion dynamics are: (1) a strong rarefaction wave
generated from the expansion of the radiation ablated plasmas
seriously decreases the ablation pressure and limits the possibility
of increasing implosion velocity; (2) hydrodynamic instabilities are
more severe than predicted causing hot spot asymmetry and material
mixing, and one of the important reasons for this is multiple shock
reflections at the fuel/hot-spot interface during the deceleration
phase.

\begin{figure}[htbp]
  \includegraphics*[height=2.4in, angle=270] {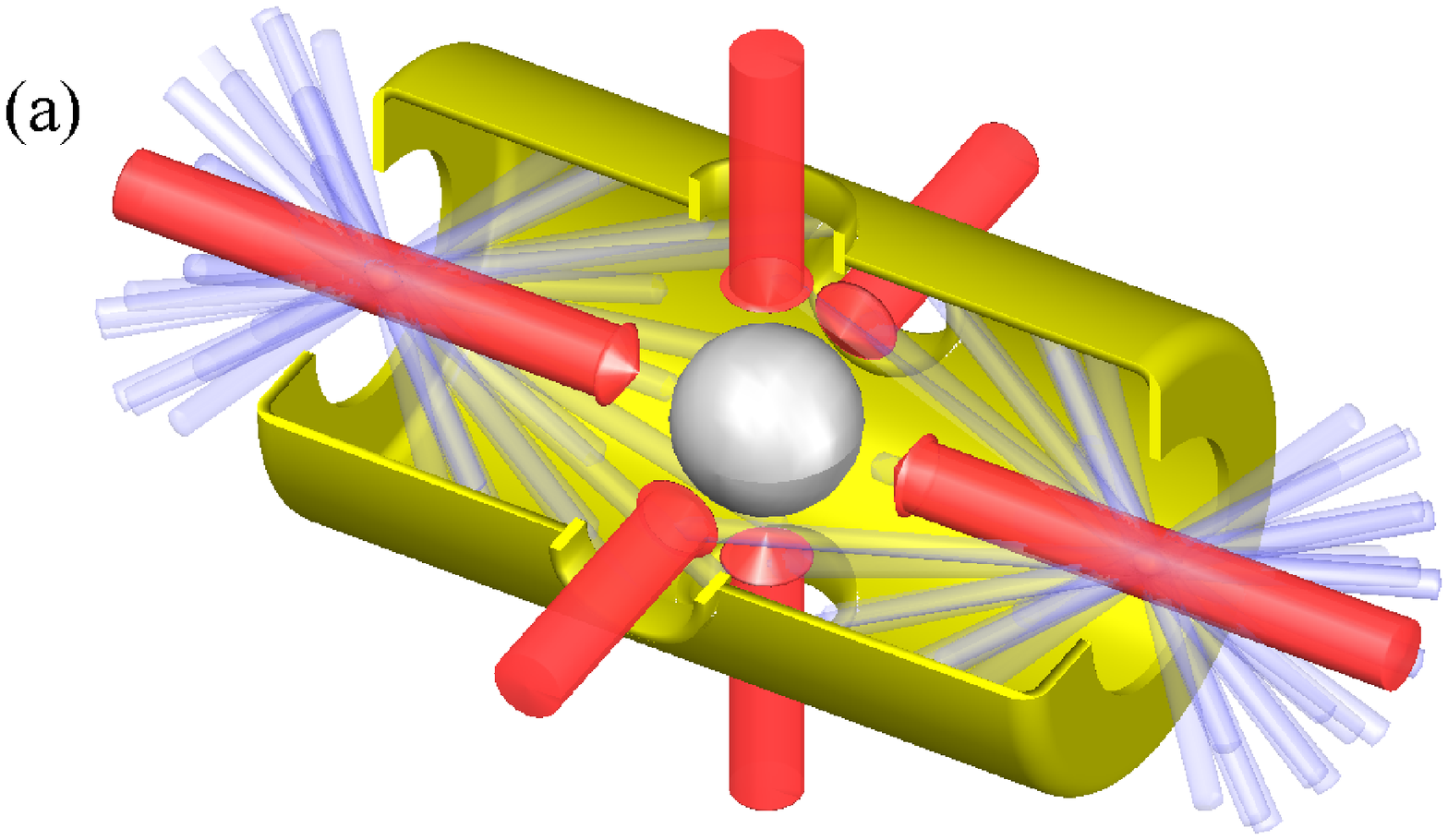} \\
  \includegraphics[width=1.6in, height=1.6in] {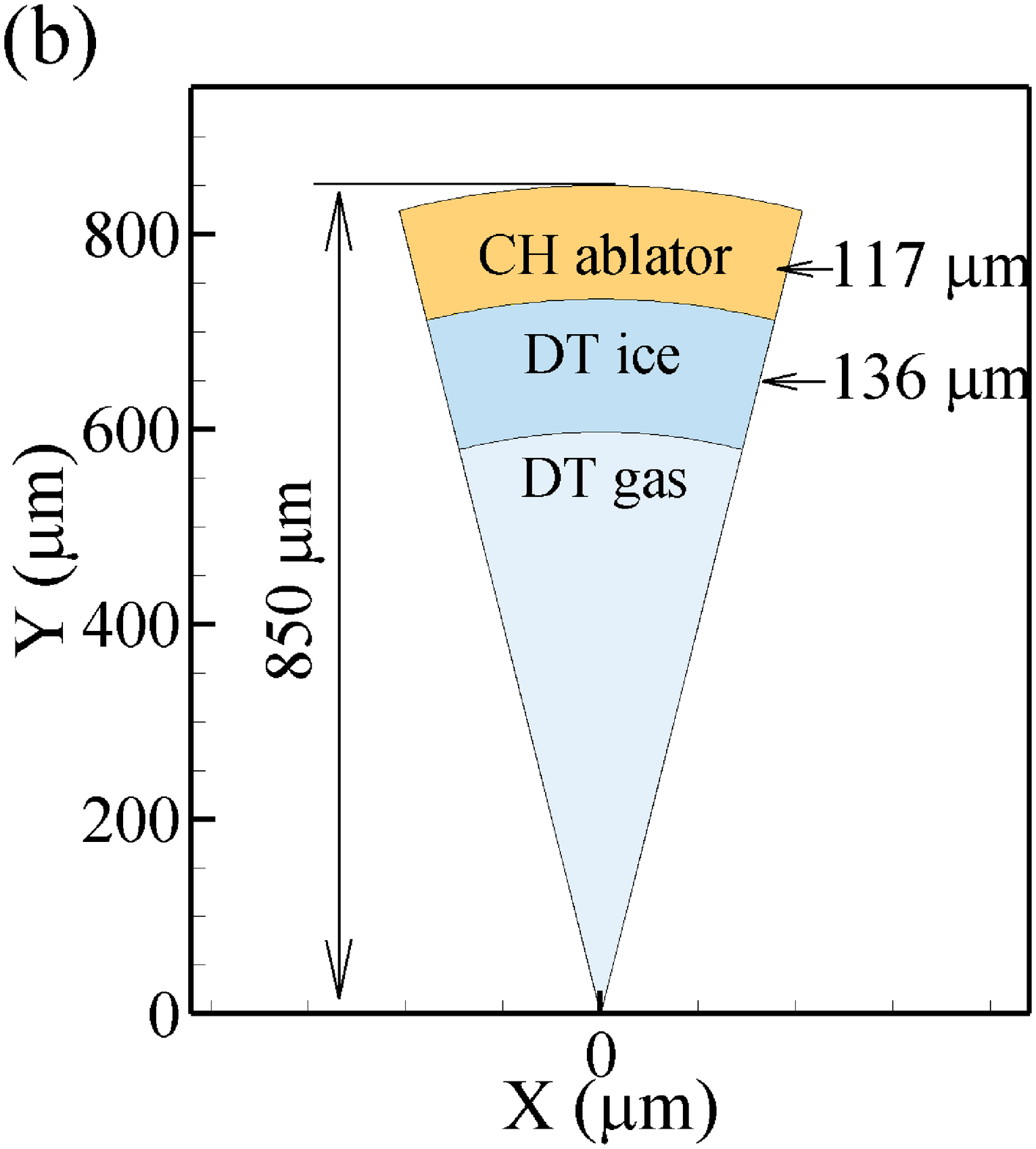}
  \includegraphics[width=1.7in, height=1.6in] {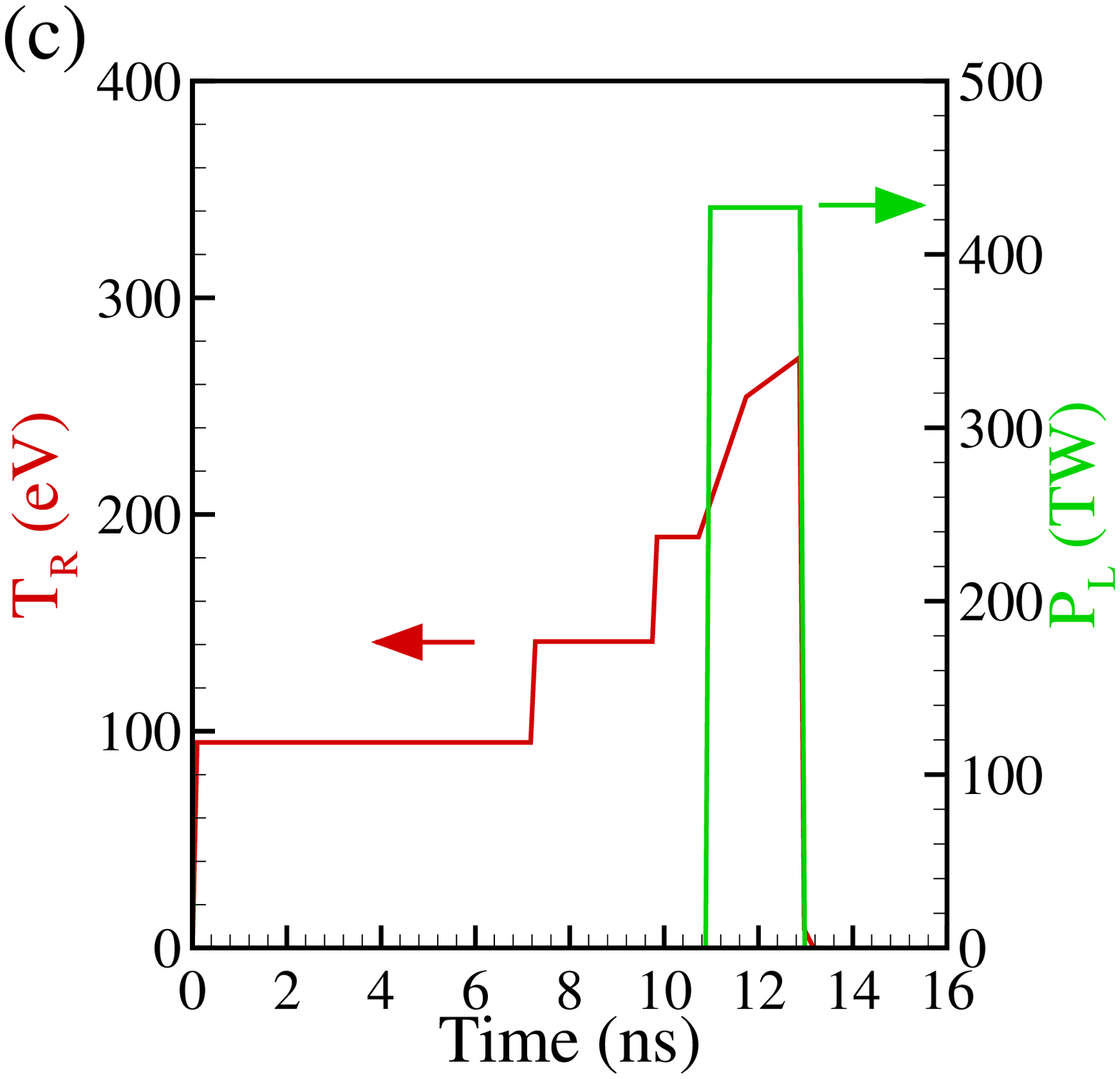} \\
  \caption{(Color online). (a) Schematic of the hybrid-drive ignition target configuration; (b) capsule
  cross-section; and (c) indirect-drive radiation temperature (red) and direct-drive laser power (green) vs time.}
  \label{Fig_1}
\end{figure}

In this Letter, to overcome the above issues, we propose a new
hybrid indirect-direct-drive ignition scheme, whose configuration is
shown in Fig. 1(a). A cryogenic ignition capsule encased at the
center of a normal cylindrical high-Z hohlraum is first compressed
symmetrically on a low adiabat by x-rays converted from
indirect-drive laser beams through two laser-entrance holes (LEHs)
at opposite ends. Then six clusters of direct laser beams are
incident straightly on the capsule through the two LEHs at the ends
and four additional symmetrical direct laser channel holes at the
waist. The hybrid indirect-direct drive accelerates and finally
ignites the capsule. This scheme is different from the previous
hybrid-drive concept \cite{Nishimura_2000} in which an initial x-ray
pulse is used for suppressing initial laser imprints by preheating,
but not during the implosion and ignition. As compared with the
recently proposed shock ignition scheme in which the compression and
ignition steps are separated and both adopt direct-drive lasers
\cite{Betti_2007, Perkins_2009, Ribeyre_2009, Schmitt_2010,
Atzeni_2012}, our hybrid-drive scheme launches direct laser beams
simultaneously with the main pulse of the indirect drive, which
creates a double-ablation-front (DAF) structure consisting of a
radiation ablation front (RAF) and an electron ablation front (EAF).
Consequently, the DAF structure results in the formation of a nearly
steady high-density plateau, which suppresses the rarefaction at the
RAF and greatly enhances the drive pressure. It is found that the
enhanced drive pressure pushes the capsule to a higher implosion
velocity and quickens the hot spot formation via inward
shock/compression waves. Meanwhile, the convergence ratio is kept at
a lower level. The hybrid-drive scheme can smooth the imprints and
drive asymmetries of direct lasers by keeping the critical surface
in the corona at a proper distance away from the capsule, and more
beneficially, hydrodynamic instabilities are greatly stabilized even
by comparison with the conventional indirect drive as in the point
design target (PT) \cite{Haan_2011} in the NIC mission.

To illustrate our new scheme, we investigate implosion dynamics of a
typical ignition capsule with an outer radius of 850 $\mu$m, about
4/5 of the PT \cite{Haan_2011} in the NIC mission. This capsule
requires a total drive laser energy of $\sim$1.35 MJ, which is also
a representative value for the PT. The cross-section of the capsule
is shown in Fig. 1(b). The CH ablator of the capsule has a 117
$\mu$m thickness with a density of 1.0 g/cm$^3$ and a mass of 0.92
mg, while the solid DT fuel layer has a 136 $\mu$m thickness with a
density of 0.25 g/cm$^3$ and a mass of 0.19 mg. The density of the
DT filling gas is 0.3 mg/cm$^3$. The total mass of this capsule is
about 1.11 mg. Figure 1(c) plots the given radiation drive
temperature and direct-drive laser power. The peak radiation
temperature is 270 eV, and the estimated indirect laser energy is
about 500 kJ by assuming a coupling efficiency of 10\% from laser to
capsule absorption. The drive pulse of the radiation temperature has
four steps at 0.0, 7.2, 9.8 and 10.7 ns. These steps create four
successive shocks whose timing follows the Munro criteria
\cite{Munro_2001}. During the rise time of the fourth step (the main
pulse), direct-drive laser beams are launched. The 0.35-$\mu$m
direct laser pulse at 425 TW has only one single step with a
duration of 2 ns and a total energy of 850 kJ. The absorbed laser
intensity near the critical surface (its radius is $\sim$1000
$\mu$m) is about 3.4 $\times$ 10$^{15}$ W/cm$^2$.

\begin{figure}[htbp]
  \includegraphics[height=2.8in, angle=270] {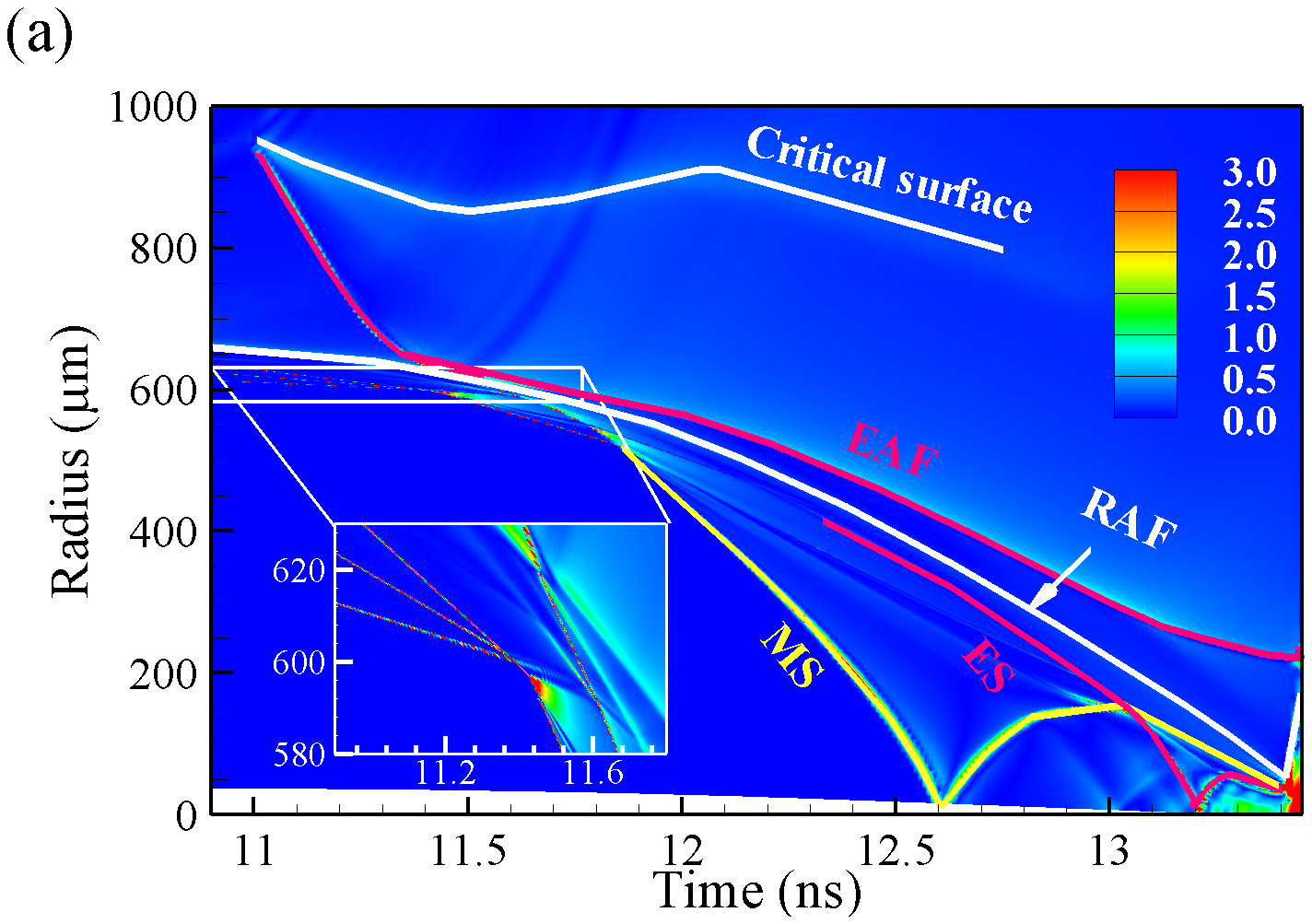} \\
  \includegraphics[height=1.4in] {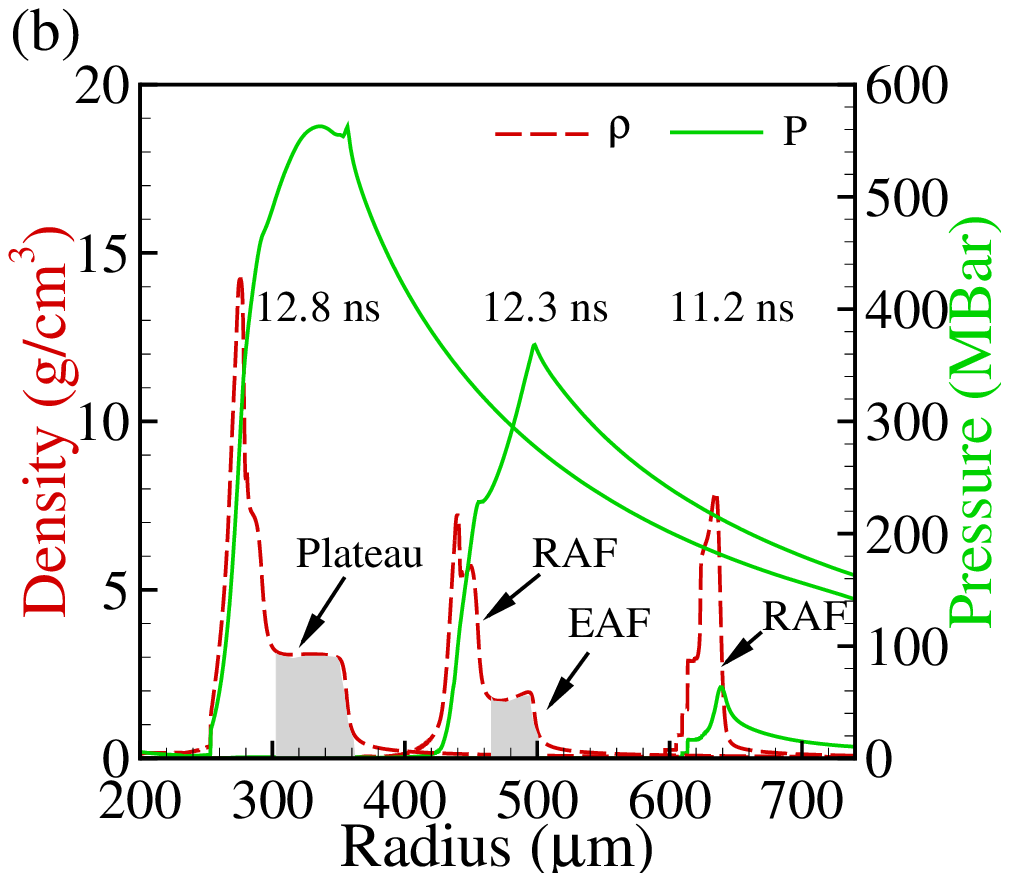}
  \includegraphics[height=1.4in] {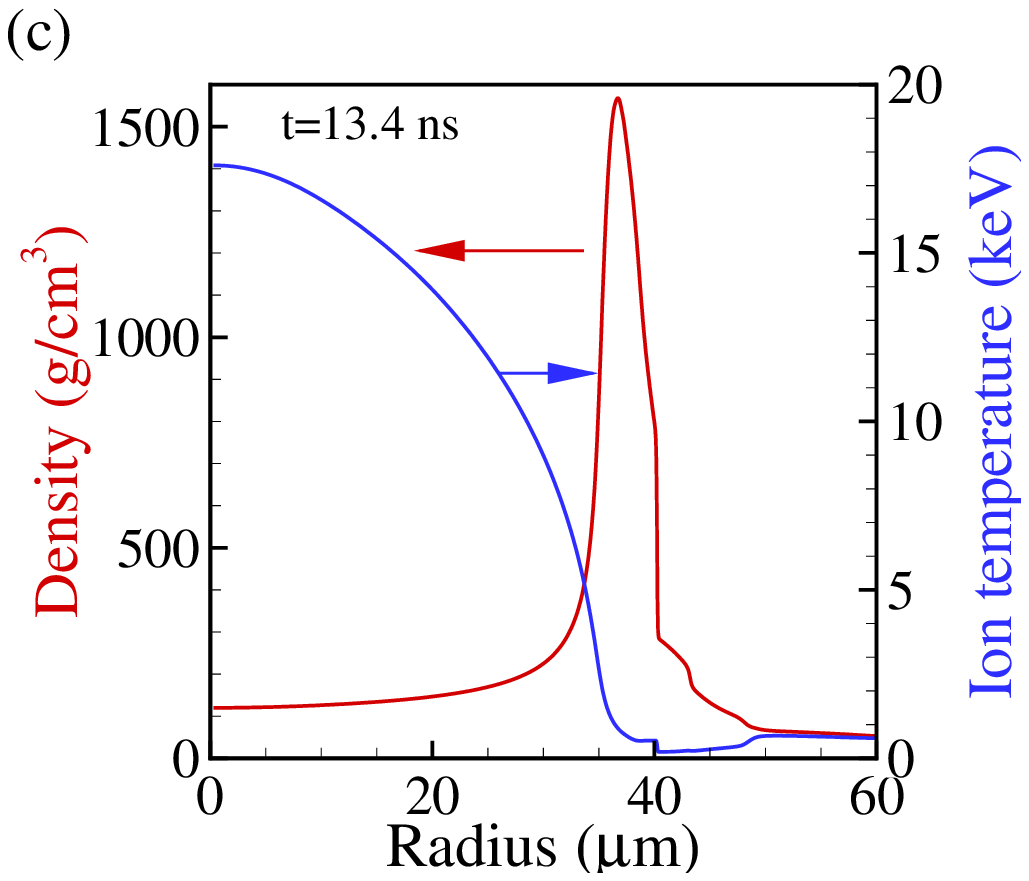} \\
  \caption{(Color online). (a) Contours of the fluid velocity gradient ($|\nabla u|$) in the space-time plane, the subpanel magnifies the four shocks of the indirect drive; (b) the density
  and pressure profiles at $t$ = 11.2 ns (with the RAF only), 12.3 ns and 12.8 ns (with both the
  RAF and EAF); and (c) the density and ion temperature profiles at stagnation $t$ = 13.4 ns.
  The EAF, RAF, MS and ES are the abbreviations of electron ablation front,
  radiation ablation front, merged shock and enhancement shock, respectively.}
  \label{Fig_2}
\end{figure}

Capsule implosion dynamics simulations are performed with the
radiation-hydrodynamics code LARED-S \cite{Pei_2007}, which is
multi-dimensional, massively parallel and Eulerian mesh based.
Multiphysics in the current simulations include laser ray tracing,
multi-group radiation diffusion (20 groups), plasma hydrodynamics,
electron and ion thermal conductions, nuclear reaction, alpha
particle transport, and the quotidian equations of state
\cite{More_1988}. In this Letter, our discussions focus only on the
implosion dynamics, while LPI involving stimulated Brillouin
scattering, stimulated Raman scattering, two-plasmon decay, etc.,
which would reduce the absorption rate of the direct-drive laser
energy and generate superthermal electrons preheating the DT fuel,
have been discussed elsewhere \cite{Depierreux_2011, Froula_2012,
Terry_2012}.

Figure 2(a) shows the implosion and ignition processes calculated by
one-dimensional (1D) simulations, using 2000 meshes with a minimum
grid size of 0.05 $\mu$m. The first three shocks of indirect drive
merge at the inner surface of the DT fuel layer at time $t$ = 11.4
ns, and the fourth shock chases them, as shown in the subpanel of
Fig. 2(a). Direct lasers launched at $t$ = 10.9 ns deposit energy in
the vicinity of the critical surface which is kept about 300 $\mu$m
away from the capsule. The electron temperature in this region
rapidly rises to a maximum of about 7 keV, and an EAF generated by
electron thermal conduction propagates towards the RAF with a
supersonic speed of $\sim$8.4 $\times$ 10$^7$ cm/s. When approaching
the RAF at about $t =$ 11.5 ns, the EAF slows down to a subsonic
speed and drives an electron thermal shock which travels into the
capsule and forms a merged shock (MS) with the previous four shocks.
After $t = 11.5$ ns, the EAF and RAF separate from each other due to
their different mass ablation rates, and hence a DAF structure is
formed. The EAF compresses the ablated rarefaction plasmas behind
the RAF like a piston, which controls the rarefaction effect from
the conventional indirect drive and results in a nearly steady
high-density plateau with a width of tens of microns, indicated by
the gray regions in Fig. 2(b). The density is enhanced from
$\sim$0.2 g/cm$^3$ at $t$ = 11.2 ns to 1.5$\sim$2.0 g/cm$^3$ at $t$
= 12.3 ns, and correspondingly the pressure at the RAF is
significantly increased from $\sim$60 MBar to $\sim$230 MBar, as
shown in Fig. 2(b). During the implosion process, the drive pressure
is further increased as the capsule converges, with an average value
of 450 MBar from $t$ = 11.5 to 13.2 ns, leading to a maximum
implosion velocity of 4.25 $\times$ 10$^7$ cm/s. Comparing with the
PT \cite{Haan_2011} in the NIC mission, at equivalent total drive
energy of $\sim$1.35 MJ, the (maximum) implosion velocity of our
hybrid-drive capsule is about 15\% higher. Meanwhile, the increasing
drive pressure produces a series of compression waves forming an
enhancement shock (ES) at about $t$ = 12.3 ns. The ES collides with
the rebounded MS near the fuel/hot-spot interface generating an
inward shock and an outward one. At about $t$ = 13.2 ns, the inward
shock arrives at the capsule center, quickly raises the hot spot
temperature and pressure, and leads to a lower convergence ratio of
25. The hot spot rapidly achieves the ignition condition soon after
the first shock reflection ($\sim$13.3 ns) at the fuel/hot-spot
interface. If the ES does not exist as in the conventional indirect
drive, there would have multiple shock reflections at the
fuel/hot-spot interface followed by severe Rayleigh-Taylor
instability (RTI) growth. Figure 2(c) shows the stagnation density
and temperature profiles at about $t$ = 13.4 ns. The corresponding
peak fuel density, hot spot areal density and hot spot average ion
temperature are 1560 g/cm$^3$, 0.59 g/cm$^2$ and 8.8 keV,
respectively. Finally, the DT fuel burns, achieving an energy yield
of 17.4 MJ and an energy gain of 13.

The drive asymmetries of the hybrid-drive ignition scheme originate
from both direct laser drive and the x-ray drive. The high-mode
direct-drive asymmetries can be smoothed by the inward supersonic
propagation of electron thermal conduction, from which the critical
surface is kept hundreds of microns away from the capsule. In this
Letter, we focus on the intrinsic low-mode direct-drive asymmetries
caused by the six limited incident directions. A three-dimensional
(3D) ray-tracing package is used to calculate the energy deposition
and evaluate the asymmetries, supposing that the laser intensity
($\sim$3.4 $\times$ 10$^{15}$ W/cm$^2$) of each cluster is uniform.
Figure 3(a) shows the distribution of absorbed laser intensities
along the polar angle $\theta$ and the azimuthal angle $\phi$
obtained at a typical time $t$ = 12.0 ns using about one million
laser rays, and the peak-to-valley ratio is 1.27. Spherical
harmonics expansion indicates that the main modes are $Y_{4,\pm 4}$
and $Y_{4,0}$ with amplitudes of $Y_{4,\pm 4}$/$Y_{0,0}$ = -2.6\%
and $Y_{4,0}$/$Y_{0,0}$ = -4.2\%, respectively. Strictly speaking,
accurate evaluation of the intrinsic low-mode drive asymmetries
needs 3D simulations which are computationally expensive. Instead,
we perform two-dimensional (2D) implosion dynamics simulations with
the $Y_{6,0}$ mode to approximate the 3D behavior, and the
equivalent amplitude matching the peak-to-valley ratio of 1.27 is
$Y_{6,0}$/$Y_{0,0}$ $\approx$ -4.5\%. Figure 3(b) plots the
corresponding 2D density contour at stagnation. The capsule seems
spherical, and the peak-to-valley amplitude at the fuel/hot-spot
interface is 0.9 $\mu$m, only 2.6\% of the hot spot radius. The
average hot spot areal density and ion temperature are close to the
1D results, and the yield over clean (the ratio of the 2D neutron
yield to the 1D yield) is almost 1.0. This means that the
deformation of the hot spot has limited effects on the ignition.
Therefore, the primary investigation indicates that the asymmetry of
the direct drive in the hybrid-drive scheme is tolerable. In
addition, an x-ray drive asymmetry caused by the four additional
direct laser channel holes is expected to be similar to the
conventional indirect drive \cite{Note_1}.

\begin{figure}[htbp]
  \includegraphics[width=1.4in]{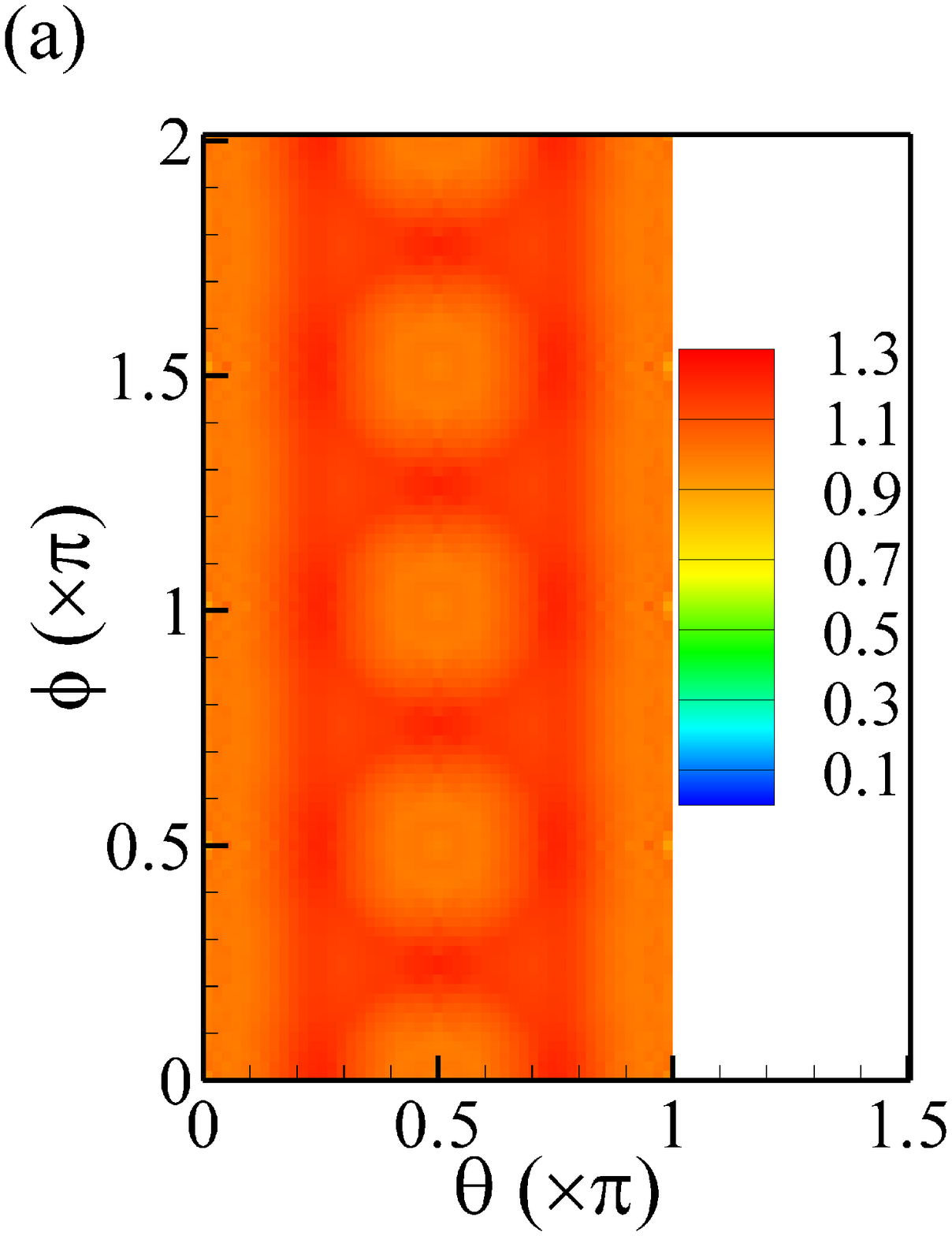}
  \includegraphics[width=1.85in]{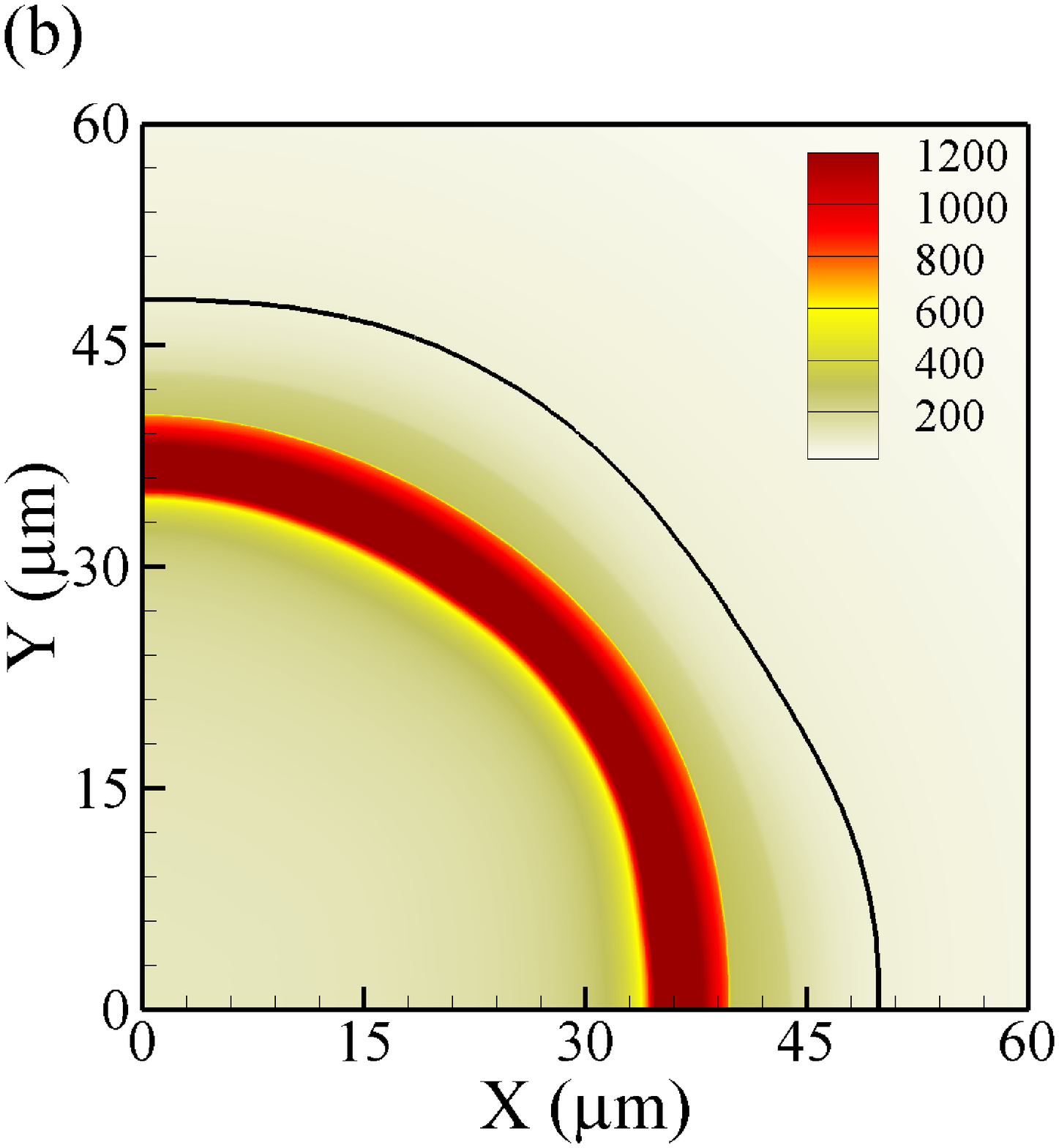} \\
  \caption{(Color online) 2D drive asymmetry of direct drive in the hybrid-drive target. (a)
  Distribution of the absorbed laser intensity normalized to the intensity
  of direct-drive laser, where $\theta$ is the polar angle
 and $\phi$ is the azimuthal angle; (b) density contour at
stagnation time for direct-drive asymmetry with $Y_{6,0}$/$Y_{0,0}$
= -4.5\%, where the black curve is the ablator/fuel
interface.}\label{Fig_3}
\end{figure}

\begin{figure}[htbp]
  \includegraphics[width=1.65in,height=1.7in] {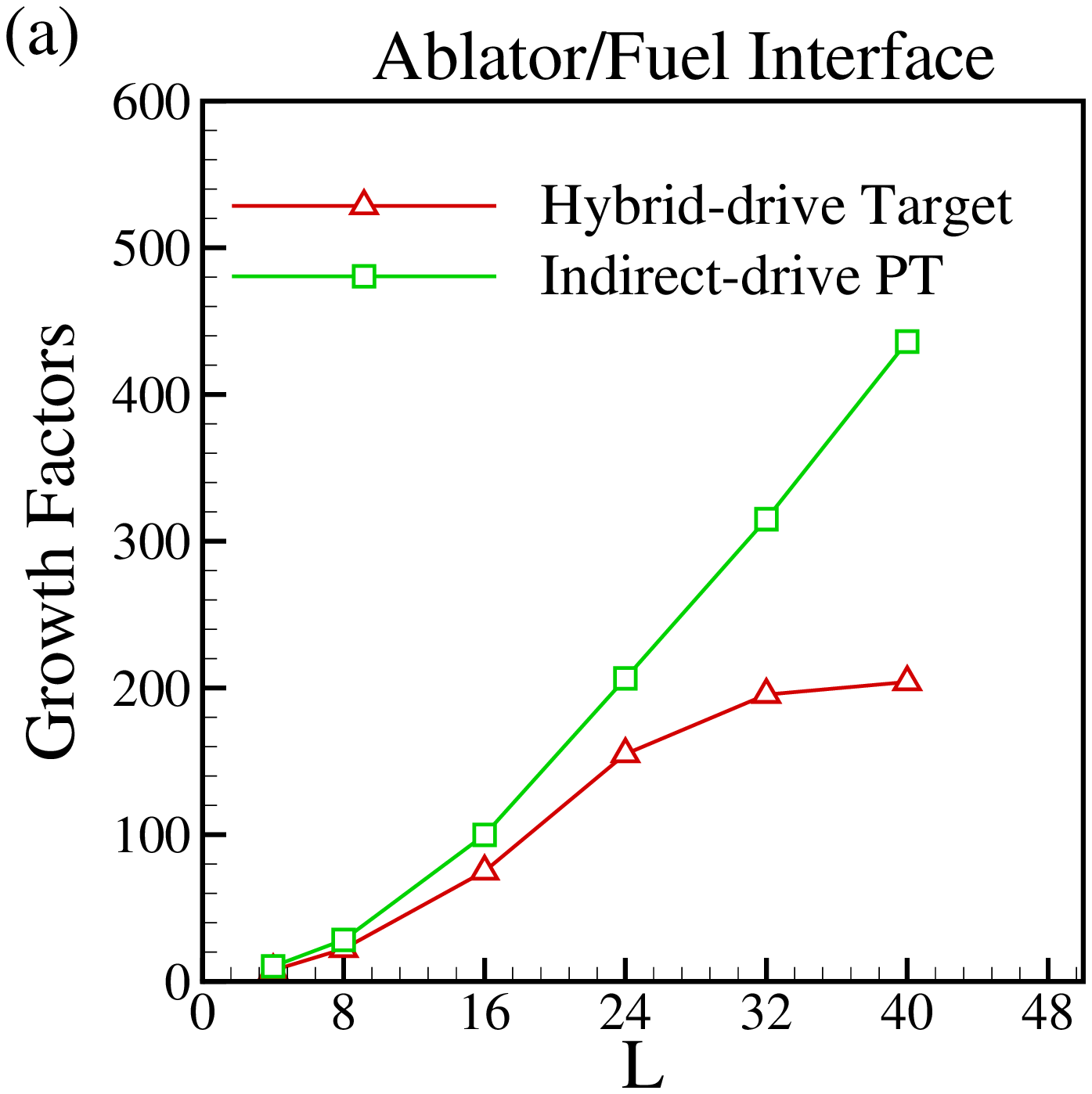} \includegraphics[width=1.65in,height=1.7in] {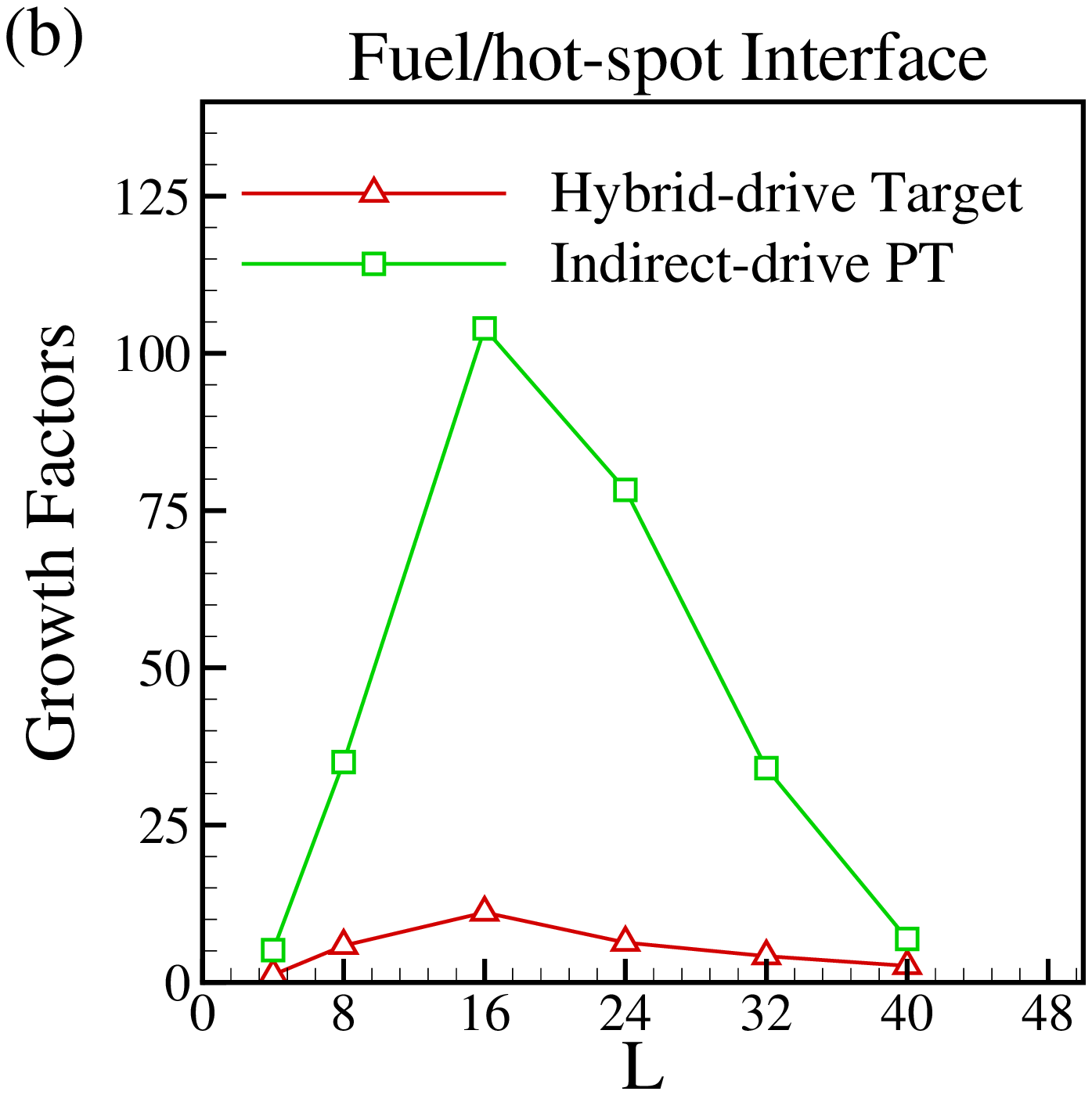} \\
  \includegraphics[width=1.55in,height=1.7in] {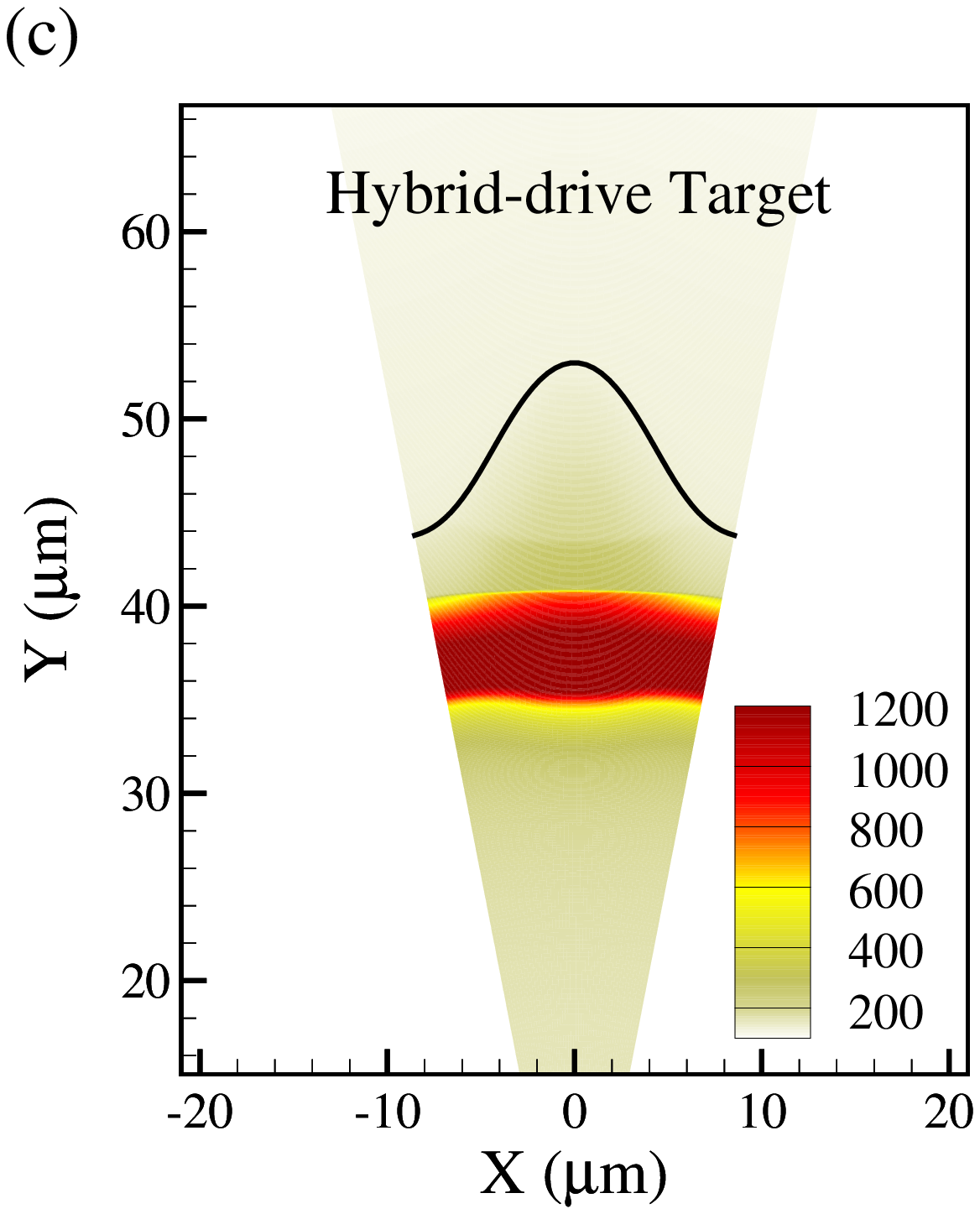} \includegraphics[width=1.55in,height=1.7in] {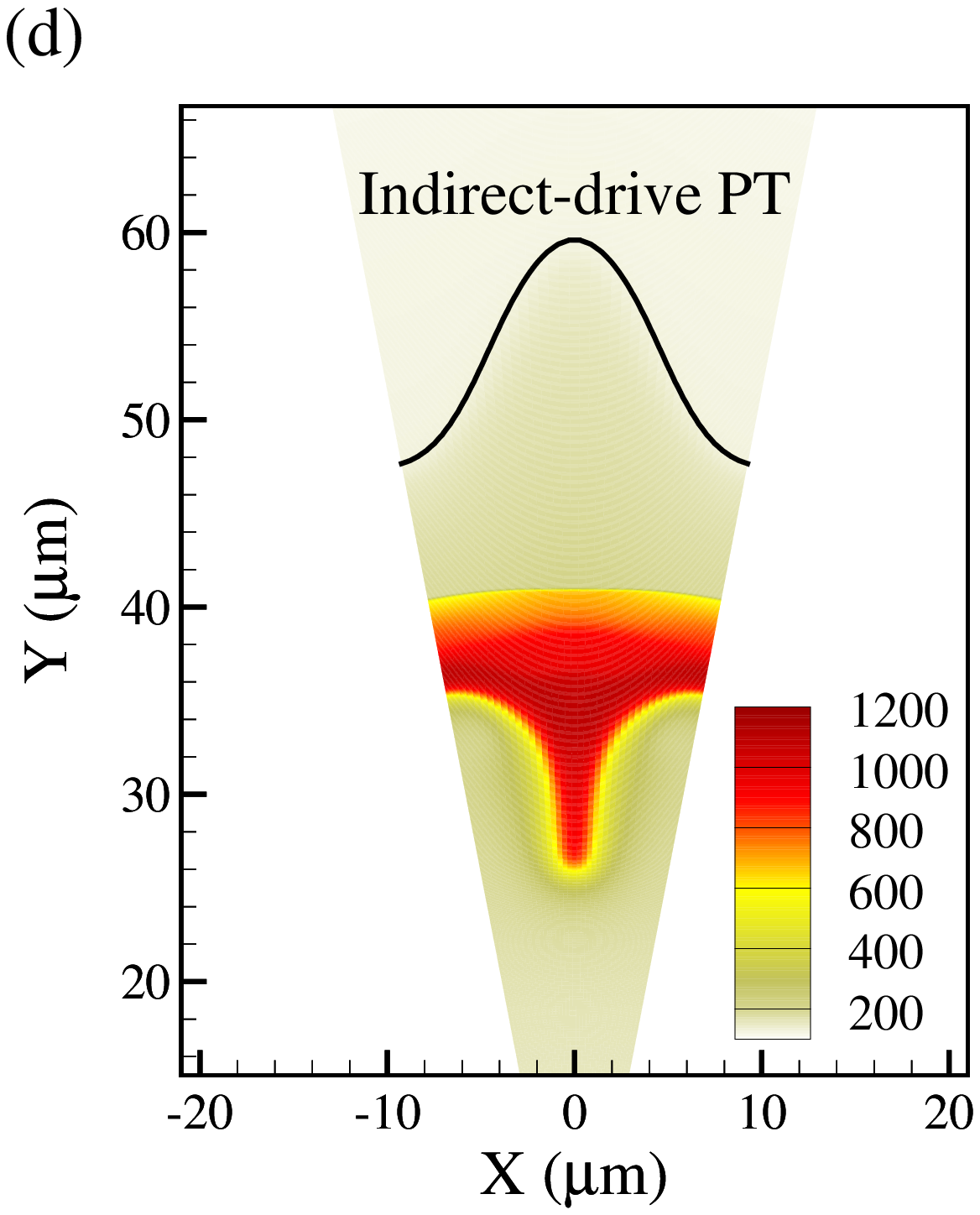} \\
  \caption{(Color online) Comparisons of the growth of hydrodynamic instabilities between the
  hybrid-drive target and the indirect-drive PT for perturbations initially seeded
  at the ablator outer surface. Growth factors
  (a) to the ablator/fuel interface and (b) to the fuel/hot-spot interface, at stagnation
  times. Density contours for a perturbation with a mode
  $L =$ 16 and an initial roughness of 350 {\AA} (c) for the hybrid-drive target, and (d) for the
  PT, at stagnation times. In (c) and (d), the black curves are the
  ablator/fuel interfaces.}\label{Fig_4}
\end{figure}

The hybrid-drive ignition features lower growth of hydrodynamic
instabilities. To demonstrate this, we compare the hybrid-drive
ignition target with an indirect-drive PT similar to that presented
in Ref. \cite{Lindl_2004}. The PT target has an outer radius of 1110
$\mu$m with a 160 $\mu$m CH ablator and 80 $\mu$m solid DT fuel. The
profile of radiation temperature for the PT is similar to that of
the hybrid-drive ignition target, with the levels of the first three
steps equal to those of the hybrid-drive target and the peak
temperature of 300 eV. The laser energy requirement is about 1.35
MJ, equivalent to that of our hybrid-drive target. 1D simulations
indicate that the implosion velocity of the PT is about 3.8 $\times$
10$^7$ cm/s, and the thermonuclear energy yield is 16 MJ. Then we
perform a series of 2D single-mode simulations with the mode number
$L$ ranging from 4 to 40. The perturbations are initially seeded on
the ablator outer surface. In order to save computation time, the
simulations are done on wedges with angles $\theta \in[\pi/2-\pi/L,
\pi/2+\pi/L]$, and 50 meshes are used in the angular direction.

First, perturbation growth to the ablator/fuel interface is reduced
in the hybrid-drive target as compared with that in the PT, as shown
in Fig. 4(a). During the acceleration stage of the capsule
implosion, i.e. $t$ $\in [11.5, 13.2]$ ns, the DAF structure reduces
the Atwood number at the RAF to an average of 0.63. The reduced
Atwood number decreases the perturbation growth, since the RTI
growth is approximated by the modified Lindl formula
$\gamma=\sqrt{{A_t k g}/{(1+A_t k L_m)}}-\beta k V_a$
\cite{Ye_2002}, where $A_t$, $k$, $g$, $L_m$, $V_a$ and $\beta$ are
the Atwood number, wave number, acceleration, minimum density
gradient scale length, ablation velocity and a constant,
respectively.

Second, perturbation growth to the fuel/hot-spot interface is
significantly reduced in the hybrid-drive target as compared with
that in the PT, as shown in Fig. 4(b). During the deceleration stage
of the PT, a merged shock reflects twice (or even more) off the
fuel/hot-spot interface leading to an early reverse of the pressure
gradient (i.e. $\nabla p \cdot \nabla \rho <0$) and causing severe
growth of RTI. However, in the hybrid-drive target, the collision of
the ES with the rebounded MS delays the reverse of the pressure
gradient, and a higher hot spot temperature also enhances the
stabilization of deceleration phase RTI. By comparison, the growth
of the $L$ = 16 mode which has maximum growth, is reduced by an
order of magnitude in the hybrid-drive target. Figures 4(c) and 4(d)
compare the density contours at stagnation times between the
hybrid-drive target and the PT, for the $L$ = 16 mode with an
initial roughness of 350 ${\AA}$. One can see that the bubble and
spike structures at the fuel/hot-spot interface are much smaller in
the hybrid-drive target than those in the PT. The lower level growth
of hydrodynamic instabilities means that the hybrid-drive target is
more robust than the indirect-drive PT, and this is essential for
realizing ignition in ICF.

In summary, we have proposed a new hybrid-drive ignition scheme
coupling both indirect drive and direct drive. Higher drive pressure
($\sim$450 Mbar) and implosion velocity ($\sim$4.3 $ \times$ $10^7$
cm/s) are obtained due to a high-density plateau between the RAF and
EAF. The ignition process is quickened and the convergence ratio is
much lower ($\sim$25). It is found that 2D simulation results with
intrinsic low-mode asymmetry of direct drive are close to the 1D
results. More importantly, the hybrid drive scheme features lower
growth of hydrodynamic instability, especially at the fuel/hot-spot
interface where the perturbation is reduced almost by an order of
magnitude when compared with the conventional indirect drive.

The authors are grateful to L. J. Perkins and Dongguo Kang for
beneficial discussions. This work has been supported by the National
ICF program of China, the National Basic Research Program of China
(No. 2013CB34100), the National High Technology Research and
Development Program of China (No. 2012AA01A303) and the National
Natural Science Foundation of China (Grant Nos. 10905006, 10935003,
11075024, 91130002, 11105016 and 11205017).

\end{CJK*}


\begin{thebibliography}{99}
\bibitem{Note_1} When the 0.35-$\mu$m lasers irradiate on the hohlraum wall,
the velocity of the critical surface is $\sim$15 $\mu$m/ns. Hence,
if we insert the bottoms of the direct laser channel holes into the
hohlraum with a depth of $\sim$150-200 $\mu$m, the overdense plasmas
ablated from the hohlraum wall would have little impact on the four
direct laser channels. This depth is only $\sim$5\% of the hohlraum
radius, thus the laser beam uniformity could be easily tuned similar
to the conventional indirect drive. Furthermore, the indirect-drive
asymmetry could be tuned by slightly adjusting the hohlraum and/or
balanced by artificial direct-drive asymmetry. In detailed target
designs, both 3D simulations and experiments are needed to tune the
indirect-drive asymmetry.
\end{thebibliography}
\end{document}